\documentclass[aip,jcp,notitlepage,twocolumn,reprint,floatfix]{revtex4-2}

\usepackage{hyperref}
\hypersetup{colorlinks=true, citecolor=blue, urlcolor=blue, linkcolor=blue}
\usepackage{xcolor}
\pdfoutput=1
\usepackage[utf8]{inputenc}
\usepackage[T1]{fontenc}
\usepackage{calc}
\usepackage{graphics,graphicx,amsfonts,amsmath,amsbsy,amssymb,color}
\usepackage{bm}
\usepackage{subfigure}
\usepackage{threeparttable}

\begin{document}

\title{Density Matrix Renormalization Group for Transcorrelated Hamiltonians: Ground and Excited States in \emph{ab initio} Systems}

\author{Ke Liao}
\email{ke.liao.whu@gmail.com}
\affiliation{
Division of Chemistry and Chemical Engineering, California Institute of Technology, Pasadena, CA 91125,
USA
}

\author{Huanchen Zhai}
\email{hczhai.ok@gmail.com}
\affiliation{
Division of Chemistry and Chemical Engineering, California Institute of Technology, Pasadena, CA 91125,
USA
}

\author{Evelin Martine Christlmaier}
\affiliation{
Max Planck Institute for Solid State Research, Heisenbergstrasse 1, 70569 Stuttgart, Germany
}

\author{Thomas Schraivogel}
\affiliation{
Max Planck Institute for Solid State Research, Heisenbergstrasse 1, 70569 Stuttgart, Germany
}

\author{Pablo L\'opez R\'ios}
\affiliation{
Max Planck Institute for Solid State Research, Heisenbergstrasse 1, 70569 Stuttgart, Germany
}

\author{Daniel Kats}
\affiliation{
Max Planck Institute for Solid State Research, Heisenbergstrasse 1, 70569 Stuttgart, Germany
}

\author{Ali Alavi}
\affiliation{
Max Planck Institute for Solid State Research, Heisenbergstrasse 1, 70569 Stuttgart, Germany
}
\affiliation{
Yusuf Hamied Department of Chemistry, University of Cambridge, Lensfield Road, Cambridge CB2 1EW, United Kingdom
}

\begin{abstract}
We present the theory of a density matrix renormalization group (DMRG) algorithm which can solve for both the ground and excited states of non-Hermitian transcorrelated Hamiltonians,
and show applications in \emph{ab initio} molecular systems. Transcorrelation (TC) accelerates the basis set convergence rate by including known physics (such as, but not limited to, the electron-electron cusp) in the Jastrow factor used for the similarity transformation. It also improves the accuracy of approximate methods 
such as coupled cluster singles and doubles (CCSD) as shown by recent studies. 
However, the non-Hermiticity of the TC Hamiltonians poses challenges for variational methods like DMRG. Imaginary-time evolution on the matrix product state (MPS) in the DMRG framework has been proposed to circumvent this problem, but this is currently limited to treating the ground state, and has lower efficiency than the time-independent DMRG (TI-DMRG) due to the need to eliminate Trotter errors. 
In this work, we show that with minimal changes to the existing TI-DMRG algorithm, namely replacing the original Davidson solver with the general Davidson solver to solve the non-Hermitian effective Hamiltonians at each site for a few low-lying right eigenstates, and following the rest of the original DMRG recipe, one can find the ground and excited states with improved efficiency compared to the original DMRG when extrapolating to the infinite bond dimension limit in the same basis set. Accelerated basis set convergence rate is also observed, as expected, within the TC framework.

\end{abstract}

\maketitle

\section{Introduction}

The density matrix renormalization group (DMRG) algorithm proposed by
Steven White\cite{white1992density,white1993density} was originally
found to be a successful method for treating one-dimensional strongly
correlated model systems.
The idea has been quickly adapted for quantum chemistry and nowadays
it is an important and reliable tool for treating static-correlation,
open-shell, and large-active-space electronic structure
problems,\cite{chan2002highly,baiardi2020density,olivares2015ab,wouters2014density,sharma2014low}
along with other approximate full configuration interaction (CI)
solvers, including full CI quantum Monte Carlo
(FCIQMC)\cite{booth2009fermion,cleland2010communications,booth2013towards,blunt2017density,guther2020neci}
and semistochastic heat-bath CI
(SHCI).\cite{holmes2016heat,sharma2017semistochastic,holmes2017excited,smith2017cheap}
The DMRG framework is also quite flexible\cite{ren2020general} and can
be combined with other ideas for studies in various fields, including
quantum dynamics,\cite{ren2022time} vibrational
spectra,\cite{baiardi2017vibrational} dynamic
response,\cite{ronca2017time,jiang2020finite} and quantum
computation.\cite{li2019electronic,lee2022there}

A difficult but important problem of using DMRG for \emph{ab initio} systems is the simultaneous treatment of both static and dynamic correlation.\cite{larsson2022chromium} Over the years, many promising ``post-DMRG'' schemes have been proposed to solve this problem.\cite{cheng2022post} Most of these schemes are based on the multi-reference theoretical framework, where DMRG, as an active space solver, is combined with other methods which are good for dynamic correlations, such as coupled cluster (CC),\cite{magoulas2021externally,lee2021externally} CI,\cite{saitow2013multireference,luo2018externally} perturbation theory,\cite{angeli2002n,roemelt2016projected,guo2016n,larsson2022matrix} canonical transformation (CT),\cite{neuscamman2010review,yanai2010multireference,neuscamman2010strongly} adiabatic connection,\cite{beran2021density} driven similarity renormalization group,\cite{evangelista2014driven,khokhlov2021toward} density functional theory,\cite{gagliardi2017multiconfiguration,sharma2019density} and transcorrelation (TC).\cite{baiardi2020transcorrelated} Although many of these methods have been shown to be useful in some benchmark systems, there is still much potential for improving their efficiency and accuracy.\cite{cheng2022post} In this work we will consider some possible improvements of the TC-DMRG approach.

In the transcorrelated method, originally proposed by Boys and Handy,
\cite{Boys1969a} the Schr\"odinger Hamiltonian is
similarity-transformed to absorb a Jastrow factor into the so-called
TC Hamiltonian, which is non-Hermitian.
Ten-no \textit{et al}.\ approached the non-Hermiticity with the
biorthogonal formulation and pioneered the work of combining TC with
second order M\o ller-Plesset perturbation theory
(MP2)~\cite{Ochi2015b} and linearized coupled cluster singles and
doubles (LCCSD).~\cite{Hino2002}
The TC method was also used in the studies of uniform electron gas by
Luo \textit{et al}.\ and others~\cite{Luo2012,Umezawa2004}.
Tsuneyuki \textit{et al}.\ applied the TC method in periodic solids
based on plane-wave basis
functions.~\cite{Sakuma2006,Ochi2012b,Ochi2016,Ochi2017}
However, all of the aforementioned work uses relatively simple
correlators, such as the F12 type,~\cite{Ten-no2004} with fixed
parameters. Such correlators can satisfy the electron-electron cusp
condition,~\cite{Kato1957} thus ameliorating the need for a large
basis set to include dynamic correlations. 

In more recent studies, the optimal choice of the correlators have been explored
in the context of FCIQMC in the
2D Hubbard model~\cite{Dobrautz2019} and CC
in 3D uniform electron gas (3D UEG).~\cite{Liao2021}
The former study reveals that by optimizing the correlator parameters
the right ground-state eigenvector can be made much more compact,
accelerating convergence of the FCIQMC calculations with respect to
walker population.
Subsequent studies combining TC and other various methods, such as
CC~\cite{Liao2021,Schraivogel2021} and
DMRG,~\cite{baiardi2020transcorrelated, Baiardi2022} exhibit
improvements in both efficiency and accuracy (at fixed excitation
level in the CC ansatz) thanks to this property.

More elaborate forms of the correlators are also explored by some
recent work,~\cite{Cohen2019, Guther2021} where the correlators are
optimized within the variational Monte Carlo (VMC) framework utilizing
the variance minimization scheme.~\cite{Guther2021}
In the present work we also adopt this strategy, but with a different
form.~\cite{Drummond2004} 
This scheme allows us to optimize the Jastrow factor for different
states under study, hence opening the door to transferring all the
advantages of TC methods demonstrated in ground state studies to
excited states.

The non-Hermiticity of the transcorrelated Hamiltonian and its proper
treatment in the DMRG framework is a less explored direction in the
field. In fact, in modern quantum chemistry and especially
DMRG-related fields, the majority of problems are associated with the
Hermitian Hamiltonian, which has many good properties. In early
studies on DMRG with the non-Hermitian Hamiltonian, Chan~et.~al. and
Mitrushenkov~et.~al. have independently proposed the general
validation of the theoretical framework of non-Hermitian
DMRG.\cite{chan2005density, mitrushenkov2003possibility} But also some
numerical instabilities were reported in
practice.\cite{chan2005density} In the recent work by Baiardi and
coworkers,\cite{baiardi2020transcorrelated} the TC-DMRG approach is
first reported and implemented using the imaginary time evolution
(ITE) approach for optimizing the states to circumvent the
non-Hermiticity problem. Very recently, they have further shown that
the ITE based TC-DMRG approach is equally applicable to \emph{ab
initio} systems for computing ground state
energies.~\cite{Baiardi2022} Some other
approaches\cite{guo2022variational} have been reported for more
general non-Hermitian Hamiltonians where eigenvalues are allowed to
have imaginary parts, but such cases are beyond the scope of the
TC framework.

In this work, we propose an alternative \emph{time independent} (TI)
approach for TC-DMRG, which is akin to the conventional TI-DMRG.
Compared to the same TI scheme proposed in
Ref.~\onlinecite{chan2005density}, our implementation is
``one-sided''. Namely, we only compute and store the right
eigenvectors and right eigen-matrix-product-states (MPS) of the TC
Hamiltonian. In addition, we only need to use real-number arithmetic.
So the overall algorithm is as efficient as the conventional Hermitian
DMRG. Although in principle it may not be possible to treat all
non-Hermitian Hamiltonians in this way, in practice we found this
simple scheme to work well with the non-Hermitian TC Hamiltonian,
and we observe no significant numerical issues across a wide range
of benchmark systems. In contrast with the ITE approach used by
Baiardi and coworkers,\cite{baiardi2020transcorrelated, Baiardi2022}
we show that our approach can be easily extended for treating both the
ground and excited states. In the remainder of the paper, we refer to
time-independent TC-DMRG and conventional DMRG as TC-DMRG and DMRG,
respectively, and to imaginary-time evolution TC-DMRG as ITE-TC-DMRG. 

The paper is structured as follows:\@ in the Theory section, we
recapitulate the main aspects of the TC framework and its
approximations as well as the basics of DMRG along with the extension
to non-Hermitian Hamiltonians. In the Results and Discussions section,
we first illustrate with numerical examples the convergence behavior
of the ground and excited state energies as a function of the number
of sweeps and how they can also be extrapolated to infinite bond
dimensions like in conventional DMRG, while yielding smaller
extrapolation errors.  We then showcase an accurate dissociation curve
of the N$_2$ molecule calculated by TC-DMRG already at the cc-pVTZ
basis-set level.  Finally we show as a simple example that accurate
first vertical excited state energy of the H${_2}$O molecule can be
obtained with the aug-cc-pVDZ basis set.

\section{Theory}

\subsection{Transcorrelation}
Transcorrelation \cite{Boys1969a,boys1969determination} is a technique
in which a similarity transformation is applied to the many-electron
Hamiltonian in order to absorb an exponential Jastrow correlation
factor $\hat \tau$,
\begin{equation}
    \begin{aligned}
        \bar{H} &= e^{-\hat{\tau}} \hat{H} e^{\hat{\tau}}\\
        & = \hat{H}+[\hat{H}, \hat{\tau}] + \frac{1}{2}[[\hat{H}, \hat{\tau}], \hat{\tau}]]\\
        & = \hat{H} - \sum_i \left(\frac{1}{2}\nabla_i^2\tau+\nabla_i\tau\cdot\nabla_i +\frac{1}{2}(\nabla_i\tau)^2\right),
    \end{aligned}
    \label{eq:hbar}
\end{equation}
where the Baker–Campbell–Hausdorff expansion terminates exactly at
second order because the correlator only depends on the electronic
positions, as will be discussed in the next subsection.  The TC
Hamiltonian is non-Hermitian due to the presence of
$\nabla_i\tau\cdot\nabla_i$ and contains additional terms involving up
to 3-body interactions arising from $\frac{1}{2}(\nabla_i\tau)^2$. 

\subsubsection{Correlator and its optimization}

In this study we use the Drummond-Towler-Needs form of the correlator
\cite{Drummond2004, LopezRios_Jastrow_2012},
\begin{equation}
    \hat{\tau} = \sum_{i>j}u(r_{ij})+\sum_{I=1}^{N_{\rm n}}\sum_{i=1}^{N}\chi_I(r_{iI})+\sum_{I=1}^{N_{\rm n}}\sum^N_{i>j}f_I(r_{iI}, r_{jI}, r_{ij}),
\end{equation}
where $i$ and $j$ run over the $N$ electrons and $I$ over the $N_{\rm
n}$ nuclei, and each of $u$, $\chi$, and $f$ are natural power
expansions of their arguments whose linear coefficients are treated as
optimizable paramteters, premultiplied by polynomial cutoff functions
to constrain the range of the correlator.
The electron-electron cusp is included in $u(r_{ij})$, and we choose
to augment the cuspless $\chi_I(r_{iI})$ term with a term enforcing
the electron-nucleus cusp $\Lambda(r_{iI})$ for consistency throughout
the calculation \cite{Haupt_TCopt_2022, Ma_cusp_2005}.

The optimization technique used in this work is discussed in detail in
Ref.\ \onlinecite{Haupt_TCopt_2022}, and here we only summarize the
key points. 
The correlator is optimized by minimization of the variance of the TC
reference energy,
\begin{equation}
  \sigma_{\rm ref}^2 =
    \frac {\langle \Phi_{\rm ref}|
                    e^{-\hat \tau} (\hat H-E_{\rm ref})^2
                    e^{\hat \tau}
           |\Phi_{\rm ref} \rangle}
          {\langle \Phi_{\rm ref}| \Phi_{\rm ref} \rangle} \;,
\end{equation}
where $|\Phi_{\rm ref}\rangle$ is typically the HF determinant, and
the reference energy is
\begin{equation}
  E_{\rm ref} = \frac {\langle \Phi_{\rm ref}| e^{-{\hat \tau}}
                       \hat H e^{\hat \tau} |\Phi_{\rm ref} \rangle} 
                      {\langle \Phi_{\rm ref}| \Phi_{\rm ref} \rangle}
              \;.
\end{equation}
The optimization is carried out in a variational Monte Carlo (VMC)
framework using correlated sampling, where a set of real-space
electronic configurations distributed according to $|\Phi_{\rm
ref}(\{{\bf r}_i\})|^2$ are generated, and then the parameters in
$\hat \tau$ are varied so as to minimize the Monte Carlo estimate of
$\sigma_{\rm ref}^2$ keeping the set of configurations fixed.
The resulting correlator is tailored to each specific system/state
under study.
In our calculations we used the \textsc{casino} code
\cite{Needs_casino_2020} to optimize Jastrow factors.

\subsubsection{Integral evaluation and approximation to the three-body
interaction}

We evaluate the required TC Hamiltonian matrix elements using the
TCHInt library, which in turn uses \textsc{pyscf}
\cite{Sun2018,Sun2020} to generate an integration grid and to evaluate
orbital values and gradients at the grid points, and a flexible
Jastrow factor implementation \cite{LopezRios_Jastrow_2012} to
evaluate Jastrow factor gradients at the grid points.
TCHInt then performs a standard grid integration to compute the matrix
elements.
We note that each grid integration operation is independent from the
rest, therefore the computation of the TC Hamiltonian is trivially
parallelizable.  We carefully check the convergence of the results
with respect to the number of grid points.

Treating the full 3-body interaction will significantly increase the computational cost of our algorithm. However, as 
shown in the study of the 3-dimensional uniform electron gas (3D UEG) using coupled cluster methods~\cite{Liao2021} as well as for \emph{ab initio} systems,~\cite{Schraivogel2021,Schraivogel2022} neglecting the generic 3-body operators while keeping the lower normal-ordered interactions  induces only minor errors compared to the full treatment. If not otherwise specified, the normal-ordering of the 3-body
operators is with respect to the Hartree-Fock vacuum. 
In order to compute the contractions in the 3-body integrals in 
a large basis set, an efficient procedure is developed, of which the full details and the application with CC methods on a large set of benchmark molecules will be reported in a following paper by some of the authors. 

\subsection{Time-independent transcorrelated DMRG}

\subsubsection{Hermitian DMRG and the variational principle}

In the conventional spin-adapted DMRG algorithm,~\cite{sharma2012spin,wouters2014chemps2,keller2016spin} we consider a set of \( K \) orthogonal basis functions \( \{ \phi_k \} \) corresponding to the spatial orbitals in the quantum chemistry language. We can then represent the DMRG wavefunction in the Hilbert space formed by the direct product of single-orbital states, as\cite{chan2016matrix}
\begin{equation}
    |\Psi\rangle = \sum_{\{n\}} \mathbf{A}[1]^{n_1} \mathbf{A}[2]^{n_2} \cdots       
        \mathbf{A}[K]^{n_K} |n_1\ n_2\ \cdots \ n_K \rangle,
\end{equation}
where each \( \mathbf{A}[k]^{n_k}\ (k = 2,\cdots, K -1) \) is an \( M \times M \) matrix, and the leftmost and rightmost matrices are \( 1 \times M \) and \( M \times 1 \) vectors, respectively. The dimension \( M \) is referred as the bond dimension. The integers \( n_k = 0, 1, 2 (k = 1, \cdots, K), \) are occupation numbers in each orbital. The DMRG wavefunction ansatz is thus called a Matrix Product State (MPS).

The optimization of the ground state energy within the MPS ansatz is based on the variational principle, which is
\begin{equation}
    E_0 = \min_{|\Psi\rangle}
        \frac{ \langle \Psi|\hat{H} | \Psi\rangle }{ \langle \Psi|\Psi\rangle },
    \label{eq:var-hermi}
\end{equation}
where \( \hat{H} \) is a Hermitian Hamiltonian and \( E_0 \) is the ground-state energy. In practice, to utilize the matrix-product structure of MPS and reuse the partially contracted intermediates, the optimization of MPS is performed using the iterative DMRG sweep algorithm, where in each iteration of a sweep, we only optimize one (or two) matrix (\( \mathbf{A}[k]^{n_k}\), for example) and keep all other matrices in the MPS constant.

The optimization problem at each orbital \( k \) can be transformed into a linear eigenvalue problem, formally written as
\begin{equation}
    \mathbf{H}[k]^{\mathrm{eff}} \mathbf{\Psi}[k]^{\mathrm{eff}}
        = E[k] \mathbf{\Psi}[k]^{\mathrm{eff}},
    \label{eq:eff-eigsh}
\end{equation}
where \( \mathbf{H}[k]^{\mathrm{eff}} \) and \( \mathbf{\Psi}[k]^{\mathrm{eff}} \) are the effective Hamiltonian ``matrix'' and the effective (ground state) wavefunction ``vector'' defined at site \( k \), respectively. \( E[k] \) is the ground state energy expectation (for the whole system) found at site \( k \). Several sweeps will be performed before the energy expectation converges. To achieve high performance, the effective Hamiltonian is never constructed explicitly. When the Hamiltonian \( \hat{H} \) is Hermitian, the matrix \( \mathbf{H}[k]^{\mathrm{eff}} \) will also be Hermitian. Therefore, we can use the standard Davidson algorithm\cite{davidson1975iterative} to solve the linear eigenvalue problem \autoref{eq:eff-eigsh}.

For technical details regarding the construction of \( \mathbf{H}[k]^{\mathrm{eff}} \) and \( \mathbf{\Psi}[k]^{\mathrm{eff}} \) from \( \hat{H} \) and \( |\Psi\rangle \), and how symmetry and parallelization can be implemented in \emph{ab initio} DMRG, we refer the readers to the review papers\cite{schollwock2005density,chan2011density,baiardi2020density} and more specialized reports.\cite{zhai2021low}

\subsubsection{DMRG with non-Hermitian TC Hamiltonians}

Instead of considering the extension of DMRG for general non-Hermitian Hamiltonians, here we will only focus on the non-Hermitian Hamiltonian that can be generated from TC. For a real-number-valued \emph{ab initio} (Hermitian) Hamiltonian and a real-number-valued transcorrelator, the TC Hamiltonian will be non-Hermitian but still real. A general real non-Hermitian Hamiltonian can in principle have complex eigenvalues, but since the transcorrelator cannot change the spectrum of the original Hamiltonian, the eigenvalues for the TC Hamiltonian (in the complete basis set limit) must also be real. Finite basis sets and the approximate treatment of 3-body terms in principle can lead to complex eigenvalues, however, in practice it would mean that the basis sets or the approximations are too rough and should not be employed. Finally, from \autoref{eq:eff-eigsh} we can easily see that when both the effective Hamiltonian and eigenvalues are real, the eigenstates (or matrices in the MPS) should also be real. This means that it is possible to implement the TC-DMRG method using only real numbers.

However, when we solve the linear effective problem \autoref{eq:eff-eigsh} using an iterative non-Hermitian Davidson solver\cite{caricato2010comparison} (instead of exact diagonalization), a set of (real) orthogonal trial vectors is constructed and the original (real) effective Hamiltonian matrix is projected into a small (real) subspace matrix. This projection will in general not preserve the spectrum of the original matrix, so the subspace matrix can in principle have complex eigenvalues and complex eigenvectors. As we use only real numbers in the non-Hermitian Davidson solver, we have to discard the imaginary parts and this can in principle create numerical and convergence problems. Fortunately, this is not a significant problem for DMRG because the eigenvectors found in one sweep iteration are transformed and used as the initial guess for starting the Davidson algorithm in a subsequent sweep iteration.~\cite{chan2002highly} The random initial guess will only affect the initial one or two sweep iterations in the first sweep for optimizing a few boundary tensors in a MPS. Since the effective space spanned by the boundary MPS tensors is typically very small, these initial Davidson processes are very cheap. As a result, in all Davidson processes in DMRG, we almost always have a very good initial guess and the Davidson can quickly converge within tens of iterations (depending on the convergence threshold). During these close-to-convergence iterations, the subspace projection is almost exact for preserving the lowest eigenvalues of the effective Hamiltonian so that the imaginary parts of the eigenvalues (and eigenvectors) for the subspace matrix can be safely discarded.

\subsubsection{The stationary principle with non-Hermitian TC Hamiltonians}

The variational principle for Hermitian Hamiltonian in the form of \autoref{eq:var-hermi} is no longer valid when \( \hat{H} \) is a non-Hermitian Hamiltonian. But for the non-Hermitian Hamiltonian, we have the following \emph{stationary principle}.\cite{kraft2016modified,Ammar2022} Consider the Rayleigh quotient, defined as a functional
\begin{equation}
    R(|\Psi_{\mathrm{L}}\rangle, |\Psi_{\mathrm{R}}\rangle) = \frac{ \langle \Psi_{\mathrm{L}}|\hat{H} | \Psi_{\mathrm{R}}\rangle }{ \langle \Psi_{\mathrm{L}}|\Psi_{\mathrm{R}}\rangle },
\end{equation}
where \( |\Psi_{\mathrm{L}}\rangle \) and \( |\Psi_{\mathrm{R}}\rangle \) are arbitrary left and right trial wavefunctions, respectively. Then the stationary principle is that only when \( |\Psi_{\mathrm{L}}\rangle \) and \( |\Psi_{\mathrm{R}}\rangle \) are respectively the left and right eigenstates of the non-Hermitian Hamiltonian \( \hat{H} \), we have
\begin{equation}
    \frac{\partial R}{\partial |\Psi_{\mathrm{L}}\rangle}
    = \frac{\partial R}{\partial |\Psi_{\mathrm{R}}\rangle} = 0
    \label{eq:var-non-hermi}
\end{equation}
and the corresponding eigenvalue \( E \) is given by the value of \( R \) at this point.
In other words, for the non-Hermitian case, the energy expectation value with the left and right trial wavefunctions \( |\Psi_\mathrm{L}\rangle \) and \( |\Psi_\mathrm{R}\rangle \) now give the stationary point of the functional, if the  left and right trial wavefunctions are simultaneously the true left and right eigenstate wavefunctions, respectively. When the Hamiltonian is Hermitian, the left and right wavefunctions will always be identical and this stationary point becomes the minimal point when the trial wavefunction is the ground state \( |\Psi_0\rangle \), so \autoref{eq:var-hermi} is a special case of \autoref{eq:var-non-hermi}.

The above stationary principle with non-Hermitian TC Hamiltonians now introduces a few implications in our TC-DMRG approach:
First, \autoref{eq:var-non-hermi} may indicate that in order to find the eigenvalues of a non-Hermitian \( \hat{H} \), we need to perform the optimization with both the left and right trial wavefunctions. However, as indicated in the work (in the EOM-CCSD context) by Caricato~et.~al.,\cite{caricato2010comparison} it is possible to perform the generalized Davidson algorithm to find the eigenvalues with only the right trial wavefunctions, and any explicit construction of the left eigenvectors can be avoided. They further showed that this ``one-sided'' approach is more efficient and numerically stable than the ``two-sided'' approaches. To be precise, in our proposed approach, both the Davidson iterations and DMRG iterations are one-sided, and only the right trial eigenvector and right MPS are stored and manipulated. This corresponds to representing the trial wavefunctions \( |\Psi_{\mathrm{L}}\rangle \) and \( |\Psi_{\mathrm{R}}\rangle \) in the same subspace, but we only update the Davidson subspace and perform DMRG renormalization for optimizing the right eigenvectors. Note that the stationary condition for \( |\Psi_{\mathrm{R}}\rangle \) can be satisfied and the correct right eigenstate can be found even when the left trial vector is kept as constant.~\cite{boys1969determination,Ammar2022} This makes the DMRG part of our non-Hermitian algorithm essentially the same as the Hermitian DMRG, and the computational cost, memory and storage requirement of the conventional DMRG can be mostly preserved. Finally, as mentioned in Ref.~\cite{baiardi2020transcorrelated}, the ability to optimize only the right eigenvector is also particularly advantageous in the TC framework since the right eigenvectors of the TC Hamiltonian can be more compact than the left ones in a CI expansion.

%Second, the monotonic convergence behavior of DMRG can be lost in TC-DMRG when the Hamiltonian %is non-Hermitian, since the termination point of the iteration is now a stationary point %rather than a minimal point. 

% K.L. The monotonic convergence behaviour of Hermitian DMRG is not guaranteed either, only the upper 
% bound to the ground state is.

Second, the non-Hermitian TC-DMRG will no longer be able to provide an upper-bound of the energy. In addition, as mentioned in Ref.~\cite{chan2005density}, the quadratic convergence of the non-Hermitian Davidson algorithm is considered to be worse than the cubic convergence of the Hermitian algorithm. This may create some problems in DMRG energy extrapolation but we have not found any severe convergence problems in practice.

\subsubsection{Non-Hermitian DMRG for excited states}

One important motivation of this study is that, within the TI-DMRG framework, it is very straightforward to extend the algorithm for finding the excited states. In this work, we compute the ground and excited states simultaneously using the state-averaged DMRG. Specifically, from \autoref{eq:eff-eigsh} we can additionally solve for a few more eigenstates
\begin{equation}
    \mathbf{H}[k]^{\mathrm{eff}} \mathbf{\Psi}_i[k]^{\mathrm{eff}}
        = E_i[k] \mathbf{\Psi}_i[k]^{\mathrm{eff}},
    \label{eq:eff-eigsh-ex}
\end{equation}
where \( E_i[k] \) are ground and excited energies with \( i = 0, 1, \cdots, N - 1 \) and \( N \) is the number of roots. The truncation of bond dimension is then based on the averaged density matrices from all computed states, and all \( N \) MPSs will share the same matrices except the one in the effective site. One can reuse any conventional state-averaged TI-DMRG code~\cite{zhai2021low} with little modification for this task.

Finally, we note that there can be some other potentially useful techniques for computing excited states. One can probably get a more accurate first excited state by doing TI-DMRG with the ground state projected out. The same projection technique can be used in ITE-DMRG for finding the excited states. However, such a procedure may introduce some additional computational cost and accuracy loss. First, for non-Hermitian Hamiltonians, the projection operator would inevitably involve both the left and right ground states, and this would require the extra computation of left eigenstates, which is completely avoided in our state-averaged approach. Second, since MPSs are just approximate representations of the true wavefunctions, the projection will not be exact. As a result, the error in left and right low-energy states will accumulate in higher excited states.

\section{Results and Discussions}

All Hermitian and non-Hermitian DMRG calculations in this work were performed using the open-source code \textsc{Block2}.\cite{zhai2021low, block2}

\subsection{Convergence behavior with sweeps in TC-DMRG}
\begin{figure}
\centering
         \centering
         \includegraphics[width=0.5\textwidth]{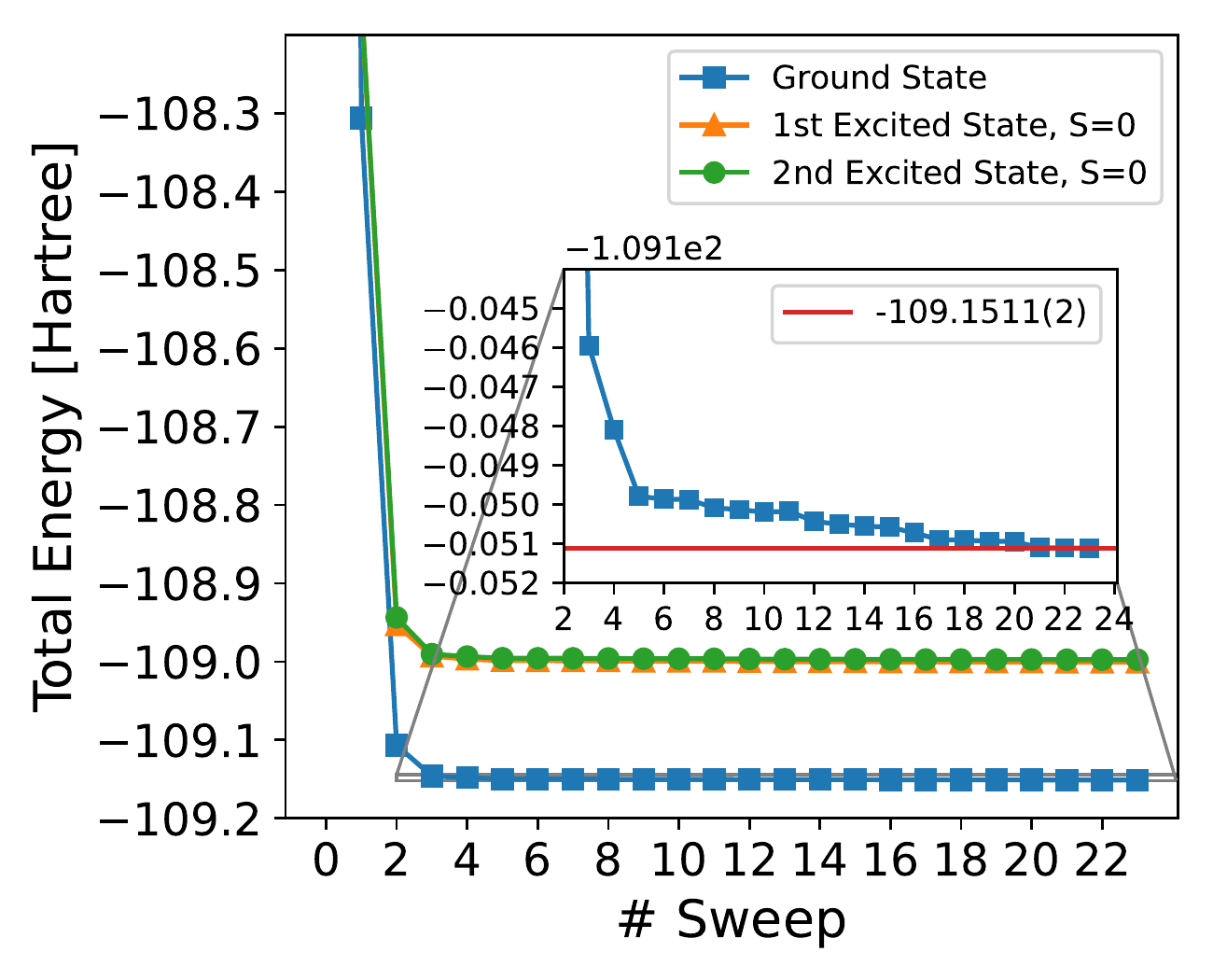}
         \caption{TC-DMRG convergence behaviors of the ground and excited state energies of the N$_{\mathrm 2}$ molecule in the cc-pVDZ basis set at 4 Bohr separation as a function of the number of sweeps and
         with increasingly large bond dimensions. Bond dimensions used are 200, 400, 600, 800, 1000 and 1200. At each bond dimension 4 sweeps are used.}
         \label{fig:conv}
\end{figure}
In \autoref{fig:conv}, we present the convergence behaviors of
the ground and excited state energies of the N$_{\mathrm 2}$ molecule,
at the stretched geometry of 4 Bohr bond length in cc-pVDZ basis set, retrieved as a function
of the number of sweeps. We increase the bond dimension 
by 200 every 4 sweeps, starting from 200 until 1200. 
The energies converge smoothly as the sweeps proceed. In practice,
we have not encountered convergence problems in all the results
presented in this work and also some other test calculations that were conducted but not included in this work, 
despite the fact that the approximations to the 3-body interactions and the projection to subspace in the general Davidson solver could result in complex eigenvalues. 
In the next subsection, we will proceed to show the extrapolation
of the energies to infinite bond dimensions, thanks to their
smooth convergence with increasingly large bond dimensions.

\subsection{Extrapolation of energies to infinite bond dimension limit}
\begin{figure}
\centering
         \centering
         \includegraphics[width=0.5\textwidth]{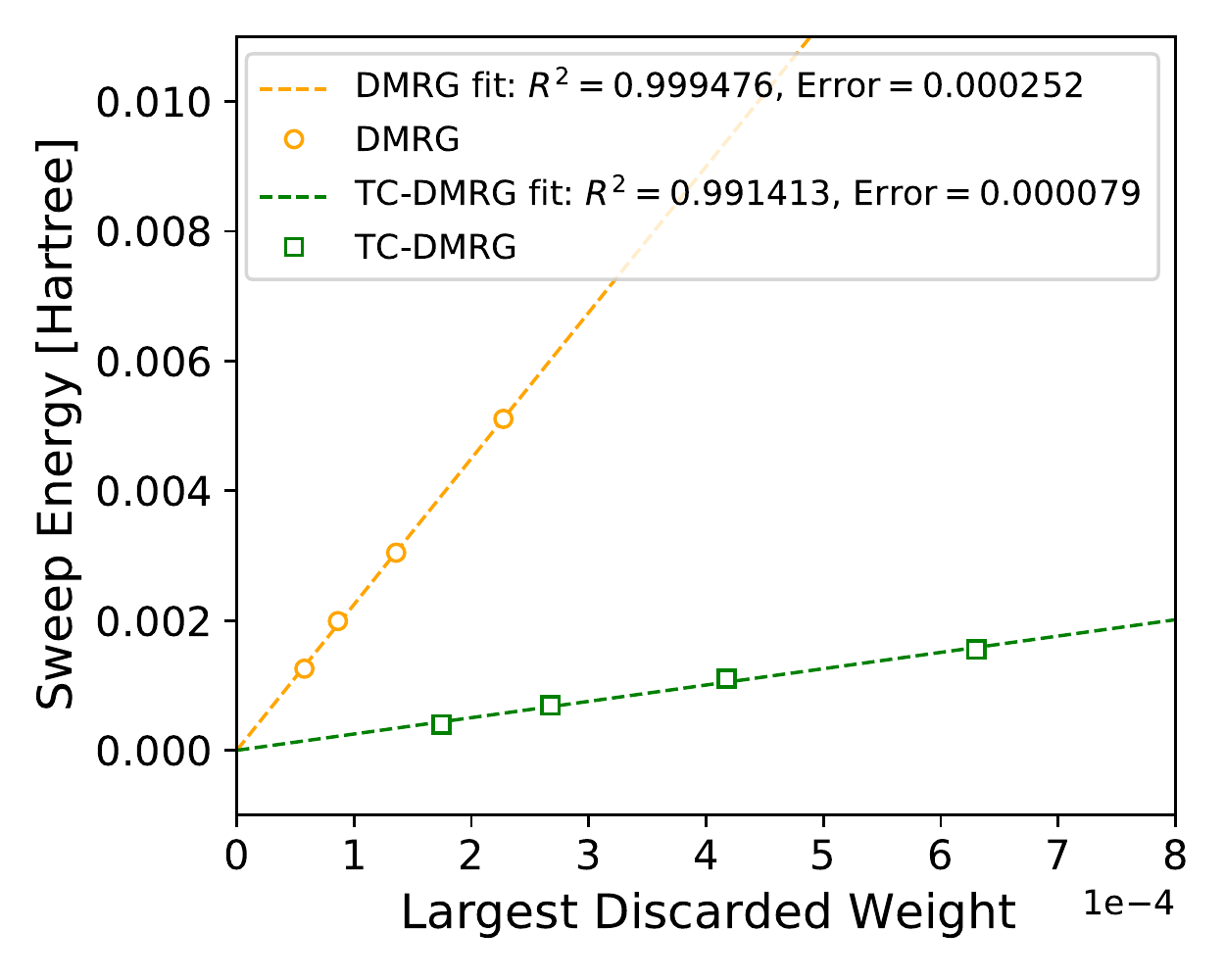}
         \caption{Extrapolations to infinite bond dimensions using linear regression for the ground state of N$_{\mathrm 2}$ calculated by DMRG and TC-DMRG, both in cc-pVDZ basis and with the respective extrapolated values subtracted.. Bond dimensions used are 200, 400, 600, 800, 1000 and 1200, where data from the smallest and largest bond dimensions are excluded. 
         The error is estimated as the absolute value of the one fifth of the difference between the extrapolated value and the data point that is closest to it.~\cite{olivares2015ab} All data points used for linear regression are obtained from the reverse schedule.~\cite{olivares2015ab}}
         \label{fig:gs}

\end{figure}

\begin{figure}
         \centering
         \includegraphics[width=0.5\textwidth]{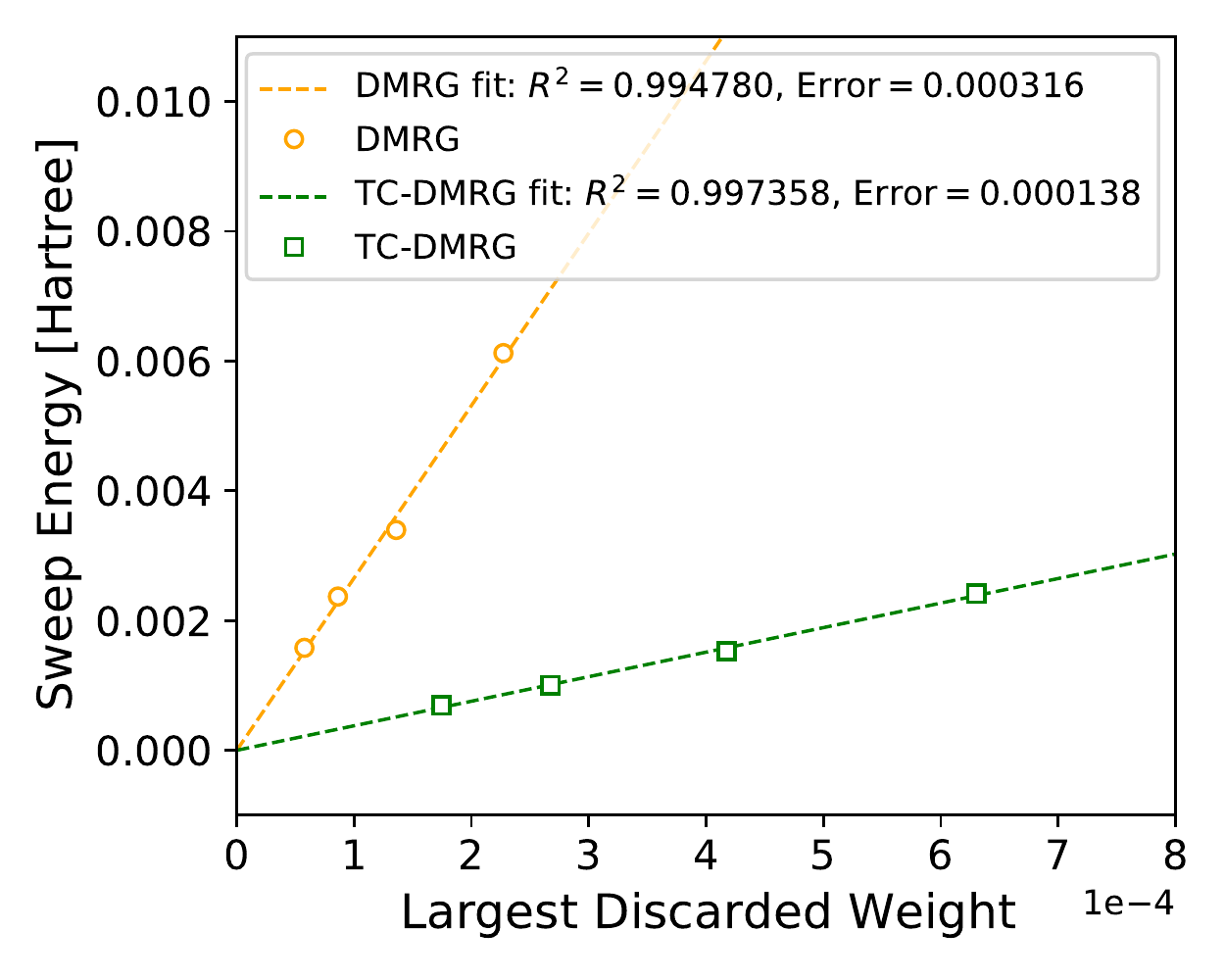}
         \caption{Extrapolations to infinite bond dimensions using linear regression for the first singlet excited state of N$_{\mathrm 2}$ calculated by DMRG and TC-DMRG, both in cc-pVDZ basis and with the respective extrapolated values subtracted. Bond dimensions used are 200, 400, 600, 800, 1000 and 1200, where data from the smallest and largest bond dimensions are excluded. The error is estimated as the absolute value of the one fifth of the difference between the extrapolated value and the data point that is closest to it.~\cite{olivares2015ab} All data points used for linear regression are obtained from the reverse schedule.~\cite{olivares2015ab}}
         \label{fig:es}
\end{figure}

In ~\autoref{fig:gs} and ~\autoref{fig:es}, a direct comparison of the linear extrapolations to infinite bond dimensions 
between the TC-DMRG and DMRG method
is shown for the ground state and first singlet excited state of the stretched N$_{\rm 2}$ molecule at 4 Bohr separation in the cc-pVDZ basis. 
In both calculations, the bond dimensions are increased
incrementally every 4 sweeps by 200, starting from
200 until 1200. For extrapolation, we follow Ref.~\cite{Eriksen2020} to exclude the data points corresponding to the smallest and largest bond dimensions and extrapolate with the largest discarded weight during the sweeps. 
Compared to DMRG, although the largest discarded weights are larger in TC-DMRG at the same bond dimensions,
the energies are closer to the extrapolated values. The extrapolation error in DMRG
is usually estimated as one fifth of the difference between the energy at the largest bond dimension (smallest largest discarded weight) used in the linear regression and the extrapolated value.~\cite{olivares2015ab}  Estimated in this way, the extrapolation errors in TC-DMRG are smaller than that in DMRG. This can be
attributed to the reduced (dynamic) correlations 
in the more compact right eigenvectors of the TC Hamiltonian compared to that of its original counterpart.


\subsection{Dissociation curve of N$_2$}
\begin{figure}
\centering
\includegraphics[width=0.5\textwidth]{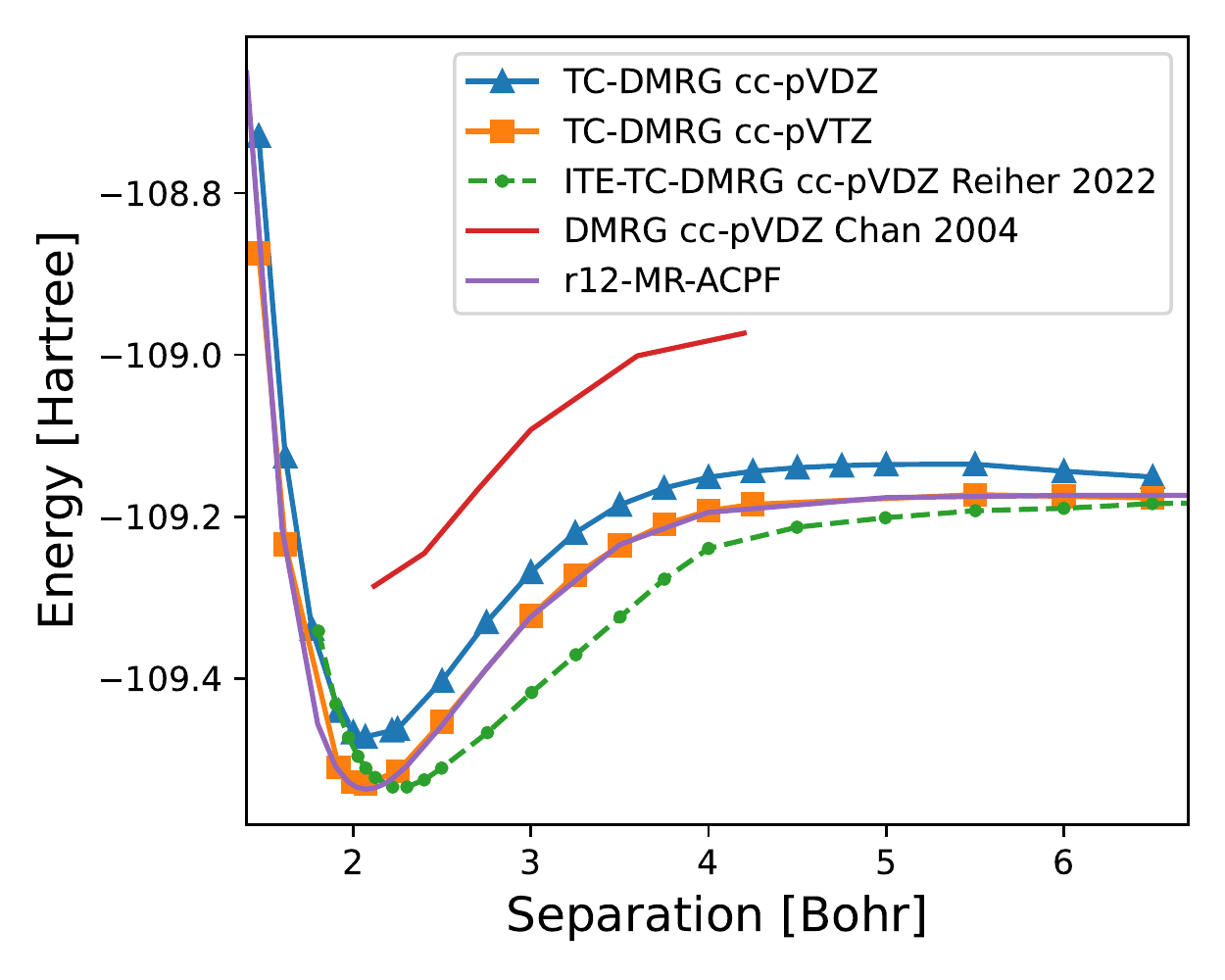}
\caption{\label{fig:n2_diss} Dissociation curve of the N$_{\mathrm 2}$ molecule. The time-independent TC-DMRG results in cc-pVDZ (blue) and cc-pVTZ (orange) basis are compared with benchmark result by
r12-MR-ACPF~\cite{Gdanitz1998} (purple). The results in the cc-pVDZ basis 
by the imaginary time evolution TC-DMRG (ITE-TC-DMRG)~\cite{Baiardi2022}
and by conventional DMRG~\cite{chan2002highly} are shown in green and red, respectively.
}
\end{figure}

The dissociation curves of the N$_{\rm 2}$ molecule calculated by various methods
are presented in \autoref{fig:n2_diss}. Compared to the DMRG at 
the same cc-pVDZ basis set,~\cite{chan2002highly} TC-DMRG produces a curve that is much closer to
the benchmark result calculated by r12-MR-ACPF.~\cite{Gdanitz1998}
When going to cc-pVTZ basis set, we get an almost 
perfect agreement with the benchmark curve. This finding also testifies the good quality of the approximation made to the 3-body interactions, not just in model systems like 3D UEG but also in an \emph{ab initio} system with strong static correlation.
For comparison, we also show 
the curve obtained by ITE-TC-DMRG from Ref.~\cite{Baiardi2022}. 
We note in passing that at large separations, the TC-DMRG curve in cc-pVDZ basis trends downwards slightly, while in cc-pVTZ basis, this trend is largely gone. 
This may hint that the current correlator
provides an imbalanced description of correlations at different separations, thus larger basis sets
are desired to remove this imbalance. The exploration of more flexible and balanced correlators will 
be left for future work, while we focus on the TC-DMRG algorithm itself in the current work. 

\subsection{Excited state of H$_2$O}
\begin{figure}
\centering
\includegraphics[width=0.5\textwidth]{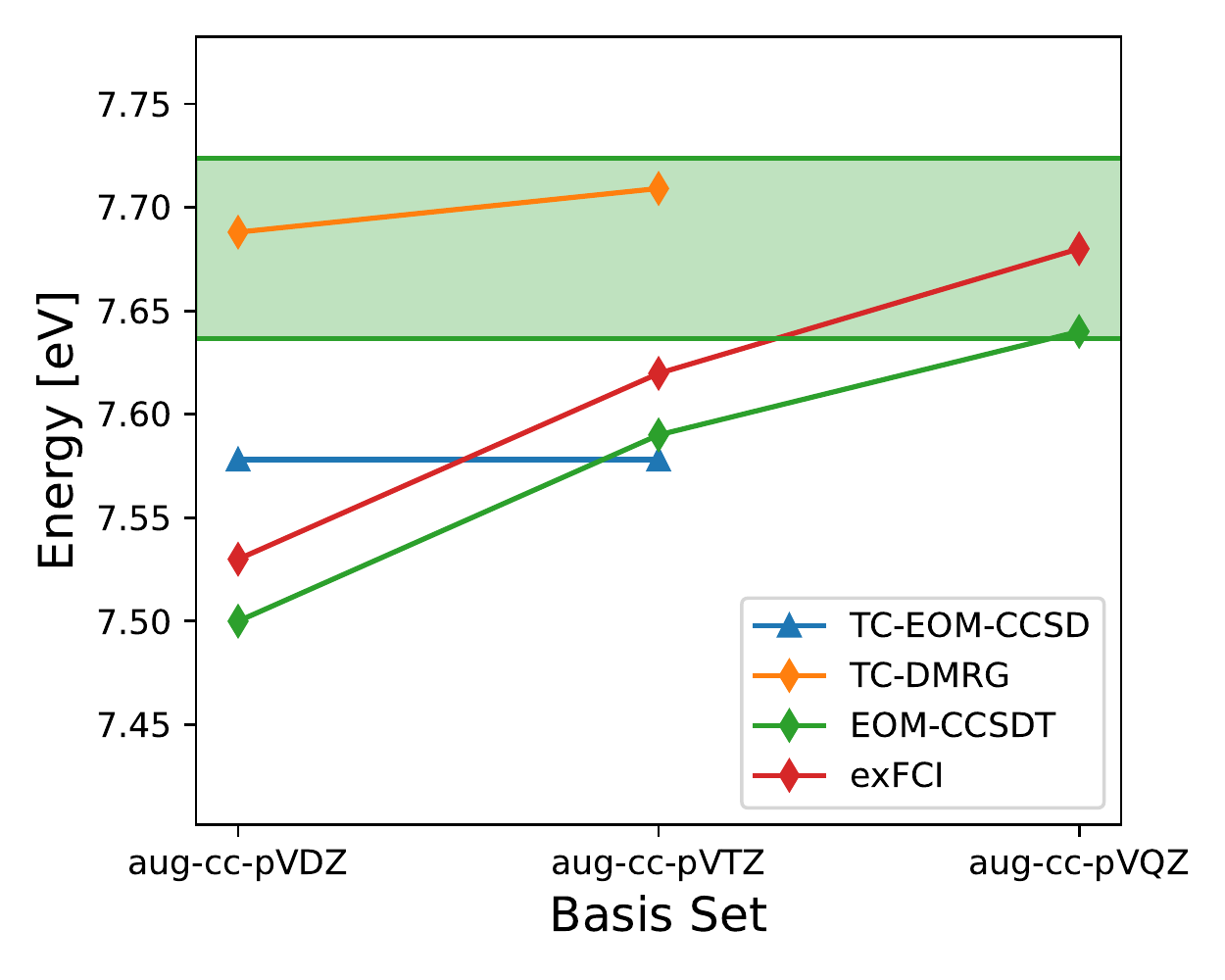}
\caption{\label{fig:h2o_excit_e} The gap between the first singly excited and the ground state of the H$_{\mathrm 2}$O molecule at
equilibrium geometry. The Jastrow factor used by TC-DMRG (orange) and TC-EOM-CCSD (blue) is the same and is optimized in the presence of a singly excited configuration state function (CSF). The benchmark results by exFCI (red) and
EOM-CCSDT (green) are taken from Ref.~\cite{Loos2018}. The green shaded area covers the chemical accuracy ($\pm$ 43 meV) around the exFCI value at aug-cc-pVQZ basis set.  
}
\end{figure}
We calculate the gap between the first singlet excited and the ground state of the water molecule at
equilibrium geometry using
increasingly large basis sets, and compare the results to the benchmark results by 
highly accurate methods (EOM-CCSDT and exFCI) found in the literature.~\cite{Loos2018}
Here we employ a state-specific Jastrow correlator that is optimized in the
presence of a singly excited configuration state function (CSF).
Already at aug-cc-pVDZ, we get with TC-DMRG a converged energy gap with 
respect to basis set between the first 
singly excited state and the
ground state, which is in agreement within the chemical accuracy ($\pm$~46 meV)
with the reference value obtained by exFCI at aug-cc-pVQZ.~\footnote{Due to the limited number of data points available, we believe no reliable extrapolation
can be performed to the complete basis set limit. So we compare the raw data at different basis sets among different methods.}

For comparison, we also plot the energy gap calculated by transcorrelated 
equation-of-motion coupled cluster singles and doubles (TC-EOM-CCSD) 
using the same TC Hamiltonian as in TC-DMRG. Although similar
convergence rate with respect to basis set is observed, TC-EOM-CCSD underestimates
the gap. 
The systematic investigation of treating excited states in molecules
within the TC framework using different methods will be explored in another coming paper.

\section{Conclusion}
In this paper, we show that small modifications in the conventional DMRG
algorithm enable it to solve non-Hermitian TC Hamiltonians of molecules for ground and excited states accurately and efficiently, thanks to the direct inclusion of dynamic correlations 
in the form of a flexible Jastrow factor. 
In our scheme, only 
the right matrix product state is stored and 
optimized by diagonalizing the effective Hamiltonians at each site iteratively, 
where the general Davidson algorithm is used in substitute of the original Davidson algorithm for obtaining a few low-lying eigenvectors. 
Both the original and the new algorithm can be understood in general as a projection scheme to find the dominant right eigenvectors in the form of an MPS, only
in the former case the Hamiltonian is Hermitian and the left and right eigenvectors are the same, hence 
follows the variational upper bound to the ground state energy. 
However, the loss of Hermiticiy is 
not a deal breaker for TC-DMRG, since small modifications to existing codes 
can make it solve for the low-lying eigenstates 
as efficient as, if not more than, the original DMRG algorithm. 
In exchange we gain the flexibility
in including dynamic correlations via the Jastrow factor and TC, which the original DMRG is poor at capturing. This makes TC-DMRG a promising tool
for treating systems where both strong static and dynamic correlations play an important role. 
What's more, we demonstrate as a preliminary study in the case of the water molecule, the
TC framework could bring the advantages it has 
for ground state to 
excited states, in that TC-DMRG and TC-EOM-CCSD achieve accelerated convergence rate with respect to the employed basis sets compared to EOM-CCSDT and exFCI. 
In this example, the Jastrow factor is optimized in the presence of the corresponding excited state CSF. This state-specific strategy would be a good choice when only a few low-lying excited states are sought accurately. 
Possible future improvements could be designing state-universal
correlators or combining TC with canonical transformation~\cite{Watson2016,Kumar2022} to reduce the dependency on the state-specific correlator when treating excited states, while retraining the existing benefits of TC. Another topic worth more careful examination is that while the current approximation to the 3-body interactions performs reasonably well in a few cases,
its general applicability still remains to be proven by more extensive numerical studies.

\begin{acknowledgments}
K.~L. and H.~Z. thank useful discussions with Garnet K.-L. Chan. This work was supported by the US Department of Energy, Office of Science, via award DE-SC0019390. This project has received funding from the European Union's Horizon 2020 research and innovation programme under Grant Agreement \#952165. The results contained in this paper reflect the authors' view only, and the EU is not responsible for any use that may be made of the information it contains.
\end{acknowledgments}

{\bf Competing Interests} The authors declare no
competing financial interests.

\bibliographystyle{achemso}
\bibliography{main}

\end{document}